\documentclass[12pt,preprint]{aastex}
\usepackage{psfig}
\usepackage{emulateapj5,apjfonts}
\usepackage[figuresright]{rotating}

\long\def\sidebyside#1#2{%
\hbox to\textwidth{\vtop{\hsize=.45\textwidth%
\parindent=0pt
\centering
 
#1\vskip1sp}\hfill\vtop{\hsize=.45\textwidth%
\parindent=0pt
\centering
#2

}}}

\newcommand{\beq}{\begin{equation}}
\newcommand{\eeq}{\end{equation}}

\def\farcd{\hbox{$.\mkern-4mu^\circ$}}

\def\earth{\hbox{$\oplus$}}
\def\arcmin{\hbox{$^\prime$}}

\def\solar{\mbox{$_{\normalsize\odot}$}}

\newcommand{\AmS}{{\protect\the\textfont2
  A\kern-.1667em\lower.5ex\hbox{M}\kern-.125emS}}
\newcommand{\lsim}{\ \raise
-2.truept\hbox{\rlap{\hbox{$\sim$}}\raise5.truept\hbox{$<$}\ }}
\newcommand{\gsim}{\ \raise
-2.truept\hbox{\rlap{\hbox{$\sim$}}\raise5.truept\hbox{$>$}\ }}
\newcommand{\simsim}{\ \raise
-1.5truept\hbox{\rlap{\hbox{$\sim$}}\raise3.5truept\hbox{$\sim$}\ }}

\slugcomment{}

\righthead{}
\lefthead{D. Gouliermis,\ W. Brandner \& Th. Henning}

\shorttitle{The Initial Mass Function of the field population in 
the LMC}
\shortauthors{Gouliermis D., Brandner W. \& Henning Th.}

\begin{document}

\title{The low-mass initial mass function of the field population 
in the Large Magellanic Cloud with {\em Hubble Space Telescope} WFPC2 
Observations} 


\author{D. Gouliermis, W. Brandner, Th. Henning}
\affil{Max-Planck-Institut f\"{u}r Astronomie, K\"{o}nigstuhl
17, D-69117 Heidelberg, Germany}
\email{dgoulier@mpia.de, brandner@mpia.de, henning@mpia.de}


\begin{abstract}

We present $V$- and $I$-equivalent HST/WFPC2 stellar photometry of an area
in the Large Magellanic Cloud (LMC), located on the western edge of the
bar of the galaxy, which accounts for the general background field of its
inner disk. The WFPC2 observations reach magnitudes as faint as $V=25$
mag, and the large sample of more than 80,000 stars allows us to determine
in detail the Present-Day Mass Function (PDMF) of the detected
main-sequence stars, which is identical to the Initial Mass Function (IMF)  
for masses $M$ {\lsim} 1 $M${\solar}. The low-mass main-sequence mass
function of the LMC field is found {\em not to have a uniform slope
throughout the observed mass range, i.e. the slope does not follow a
single power law}. This slope changes at about 1 $M${\solar} to become
more shallow for stars with smaller masses down to the lowest observed
mass of $\sim$ 0.7 $M${\solar}, {\em giving clear indications of
flattening for even smaller masses}. We verified statistically that for
stars with $M$ \lsim\ 1 $M${\solar} the IMF has a slope $\Gamma$ around
$-2$, with an indicative slope $\Gamma \simeq -1.4$ for 0.7 \lsim\
$M/M${\solar} \lsim\ 0.9, while for more massive stars the main-sequence
mass function becomes much steeper with $\Gamma \simeq -5$. The
main-sequence luminosity function (LF) of the observed field is in very
good agreement with the Galactic LF as it was previously found. Taking
into account several assumptions concerning evolutionary effects, which
should have changed through time the stellar content of the observed
field, we reconstruct qualitatively its IMF for the whole observed mass
range (0.7 \lsim\ $M/M${\solar} \lsim\ 2.3) and we find that the number of
observed evolved stars is not large enough to have affected significantly
the form of the IMF, which thus is found almost identical to the observed
PDMF.

\end{abstract}

\keywords{Magellanic Clouds --- galaxies: stellar content ---
color-magnitude diagrams --- stars: evolution --- stars: luminosity
function, mass function}

\section{Introduction}

The stellar {\em Initial Mass Function} (IMF) of a stellar system is a
quantity, which accounts for the distribution of the masses of stars
assumed to be physically related to each other (as members of the system).
This implies that this function is a physical quantity directly linked to
the formation of the specific stellar system. Authors refer to the space
outside the areas covered by stellar systems, or other stellar structures
(such as complexes or aggregates) in a galaxy, as the {\em field} of this
galaxy. The IMF in clustered regions in our galaxy, is found to be about
the same as in the solar neighborhood, and also roughly the same as the
summed IMF in whole galaxies. So, it is reasonable to assume that most
stars form in clusters and not in the general field and that stars in a
field region are not necessarily physically related to each other. Hence,
in the case of the stellar mass function of the field of a galaxy, we do
not necessarily measure a physical quantity of a uniform sample of stars
(e.g. with common origin), but rather a distribution of a random sample of
stars found in this area under different physical or statistical
circumstances (e.g. evaporation from star clusters, dissipation of open
clusters, etc).

Although the IMF appears relatively uniform when averaged over whole
clusters or large regions of galaxies (Chabrier 2003), observed local
spatial variations suggest that they may have statistical origins (e.g.
Elmegreen 1999) or that there may be different physical processes working
in different mass regimes (e.g. Elmegreen 2004). Indeed, the measured
Galactic IMF shows variations of its slope from one region to the other,
which could not only reflect sampling limitations in the observations, but
they may also be the result of physical differences, or purely statistical
in nature. Scalo (1998) notes that while the average slope of the IMF at
intermediate to high mass is about the value found originally by Salpeter
(1955), i.e., $\Gamma = -1.35$ for stellar counting in equal logarithmic
intervals, the slopes for individual regions vary by $\pm 0.5$. It is
still unknown if these variations result from physical differences in the
intrinsic IMFs for each region, or from statistical fluctuations around a
universal IMF. For example in the case of the Orion association, Brown
(1998), using Hipparcos data to determine membership, and photometry to
determine masses, found an IMF slope of $\Gamma \simeq -1.8$, while a
slope fairly consistent with Salpeter's is found by Massey and
collaborators for high mass stars in most associations in the LMC and the
Milky Way, using spectroscopy to determine masses (e.g. Massey 1998).

\begin{figure*}[t!]
\centerline{\hbox{
\psfig{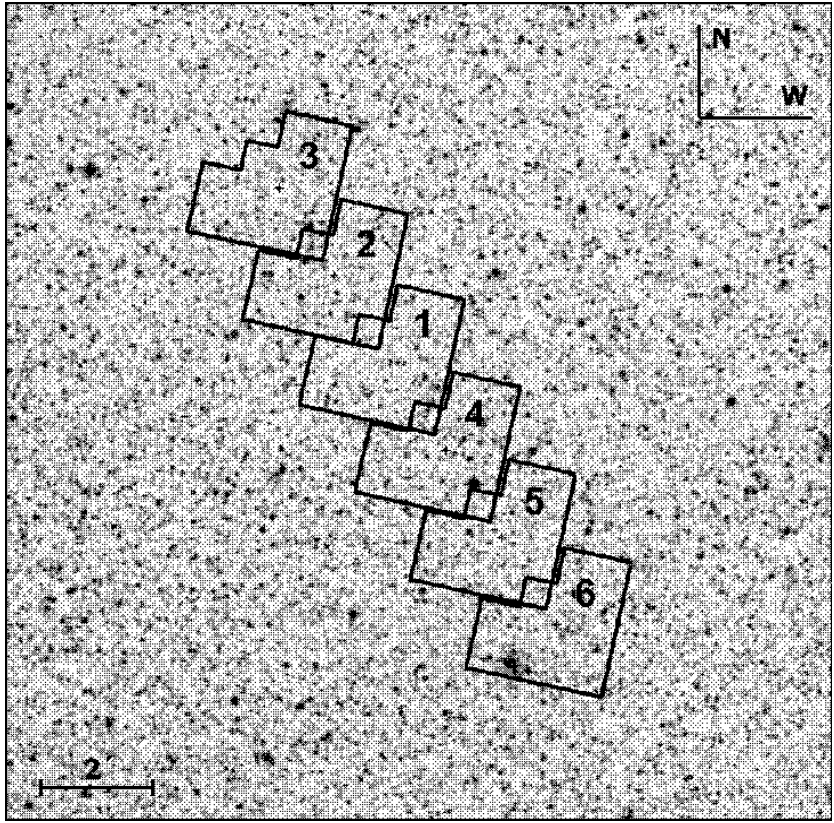}
}}
\caption{The HST/WFPC2 pointings of all six observed datasets. The
pointings are overlayed on a 15\arcmin $\times$ 15\arcmin\ chart of the
general area extracted from the SuperCOSMOS Sky Survey. The number of the
corresponding dataset is given for every pointing.}
\label{fig_map}
\end{figure*}

The Galactic IMF is found to be approximately a power law for
intermediate- to high-mass stars, but it becomes flat at the low-mass
regime (Reid 1998), down to the limit of detection, which is around 0.1
$M${\solar} or lower (Lada et al. 1998; Scalo 1986, 1998). The stellar
mass at the threshold of the flat part is considered as the thermal Jeans
mass in the cloud core, $M_{\rm J0}$ (see review by Chabrier 2003). This
mass seems to be higher in star-burst regions (e.g. Scalo 1990; Zinnecker
1996; Leitherer 1998), giving a larger proportion of high-mass stars
compared to the solar neighborhood. The proportion of high- and low-mass
stars seems to differ for Galactic cluster and field populations as well,
in the sense that the slope of the local field star IMF is steeper than
the slope of the cluster IMF (e.g. Elmegreen 1997, 1998), which is the
case for example for intermediate-mass stars in the solar neighborhood
(Scalo 1986). The same observation has been made for the remote field of
the LMC for stars in the high-mass (Massey et al. 1995) and the low-mass
regime (Gouliermis et al. 2005).

Observed variations from one region of the Galaxy to the other in the
numbers of low-mass stars and brown dwarfs over the number of
intermediate-mass stars affect the corresponding IMF, which seems to
depend on the position. In Taurus (Luhman 2000; Brice\~no et al. 2002) and
in IC 348 (Preibisch et al. 2003; Muench et al. 2003; Luhman et al. 2003)
the ratio of sub-stellar over stellar objects has been found to be almost
half of the value found in the Orion Trapezium cluster (Hillenbrand \&
Carpenter 2000; Luhman et al. 2000;  Muench et al. 2002), in Pleiades
(Bouvier et al. 1998; Luhman et al. 2000), in M35 (Barrado y Navascu\'es
et al. 2001), and the Galactic field (Reid et al. 1999). Suggested
explanations for the low-mass IMF variations include stochasticity in the
ages and ejection rates of proto-stars from dense clusters (Reipurth \&
Clarke 2001; Bate, Bonnell \& Bromm 2002; Kroupa \& Bouvier 2003),
differences in the photoevaporation rate from high-mass neighboring stars
(Preibisch et al. 2003; Kroupa \& Bouvier 2003), or differences in the
initial conditions of the turbulence (Delgado-Donate et al. 2004). They
have been also attributed to a dependence of the Jeans mass on column
density (Brice\~no et al. 2002) or Mach number (Padoan \& Nordlund 2002),
and they may also be affected by variations in the binary fraction (Malkov
\& Zinnecker 2001). Variations in the IMF have been observed also between
different mass ranges within the same region. For example Gouliermis et
al. (2005) recently report that the slope of the mass function of a region
of a stellar association (LH 52) and a field at the periphery of another
association (LH 55 field) in the LMC shows a definite change at about 2
$M${\solar} and it becomes more shallow for higher masses. This is more
prominent at the location of the star forming association, which is
characterized by a higher number of more massive stars than the field.

\begin{table*}[!t]
\begin{center}
\caption{Log of the observations. Dataset names refer to HST archive 
catalog. \label{tab1}}
\begin{tabular*}{\textwidth}[]{@{\extracolsep{\fill}}llcccccc}
\tableline
Set    & Data Set &        &      & Exposure Time &  R.A.     &Decl.  \\
number & Name     & Filter & Band & (s)           & (J2000.0) &(J2000.0) \\
\tableline
Set 1 &U63S010& F555W &WFPC2 $V$ &4 $\times$ ~500&05:01:56&$-$68:37:20 \\
      &       & F814W &WFPC2 $I$ &3 $\times$ ~700&        & \\
\tableline
Set 2 &U63S020& F555W &WFPC2 $V$ &4 $\times$ ~500&05:02:08&$-$68:35:47\\
      &       & F814W &WFPC2 $I$ &3 $\times$ ~700& & \\
\tableline
Set 3 &U63S030& F555W &WFPC2 $V$ &4 $\times$ ~500&05:02:20 &$-$68:34:14\\
      &       & F814W &WFPC2 $I$ &3 $\times$ ~700& & \\
\tableline
Set 4 &U63S040& F555W &WFPC2 $V$ &4 $\times$ ~500&05:01:44 &$-$68:38:53\\
      &       & F814W &WFPC2 $I$ &3 $\times$ ~700& &\\
\tableline
Set 5 &U63S050& F555W &WFPC2 $V$ &4 $\times$ ~500&05:01:33 &$-$68:40:26\\
      &       & F814W &WFPC2 $I$ &3 $\times$ ~700& & \\
\tableline
Set 6 &U63S060& F555W &WFPC2 $V$ &4 $\times$ ~500&05:01:21 &$-$68:41:59 \\
      &       & F814W &WFPC2 $I$ &3 $\times$ ~700& & \\
\tableline
\tableline
\end{tabular*}
\end{center}
\end{table*}

Reported variations in the IMF for high mass stars within the same stellar
system introduce the phenomenon known as primordial mass segregation,
according to which massive stars favor the young cluster cores for birth
(e.g. Le Duigou \& Kn\"odlseder 2002; Stolte et al. 2002;  Sirianni et al.
2002; Muench et al. 2003; Gouliermis et al. 2004; Lyo et al. 2004), while
the clusters have not yet the time to be mass segregated by dynamical
processes (Bonnell \& Davies 1998). This preference of massive stars
produce a flattening of the IMF. In a segregated cluster the slope of the
massive IMF can be very flat, $\Gamma\sim0$, while at the cluster
envelopes it can be as steep as $\Gamma\sim-2.5$ (de Grijs et al. 2002a).  
The mechanisms of accretion of peripheral gas (see e.g. Zinnecker
1982; Myers 2000; Larson 2002; Bonnell et al. 2001, 2004; Basu \& Jones
2004) and coalescence of other proto-stars in dense cluster cores
(Zinnecker 1986; Larson 1990; Price \& Podsiadlowski 1995; Stahler et al.
2000; Shadmehri 2004), through which high-mass stars can grow by a much
larger factor than low-mass stars, make the IMF depend on environment.
Coalescence after accretion drag (Bonnell et al. 1998) or after
accretion-induced cloud core contraction (Bonnell \& Bate 2002) also seem
likely in dense clusters in view of various simulations (Klessen 2001;
Bate et al. 2003; Bonnell et al. 2003; Gammie et al. 2003; Li et al.
2004). Furthermore, considering that the confinement of stellar winds 
and ionization during the collapse phase of massive proto-stars (Garay \& 
Lizano 1999; Yorke \& Sonnhalter 2002; Churchwell 2002;  McKee \& Tan 
2003) happens mostly in dense cloud cores, one may expect the massive star 
formation to be locked to these dense regions. Hence, the flattening of 
the IMF in dense cluster cores may be partly explained by these 
mechanisms, along with dynamical effects (Giersz \& Heggie 1996; Gerhard 
2000; Kroupa et al. 2001; Portegies-Zwart et al. 2004).

The IMF in the LMC appears to be steep in areas of the general field, away
from any stellar system. Massey et al. (1995) found that stars in the LMC
field have an IMF slope $\Gamma\sim-4$, the same value as for the Milky
Way field (see also Massey 2002). The massive stars inside known Lucke \&
Hodge (1970) OB associations in the LMC are found to have
$\Gamma=-1.08\pm0.2$, whereas the dispersed massive stars outside the
associations have a bit steeper IMFs with $\Gamma=-1.74\pm0.3$ (Hill et
al. 1994). This result is in line with Parker et al. (1998) who found that
stars not located in {\sc Hii} regions (they refer to them as field stars)  
have an IMF with $\Gamma\sim-1.8\pm0.09$, while the IMF of stars located
in {\sc Hii} regions (related to known stellar associations) exhibit
possibly three slopes: $\Gamma = -1.0$, $-1.6$, and $-2.0$. They
interprete this variability of $\Gamma$ as the result of differing star
formation processes.


\begin{figure*}[t!]
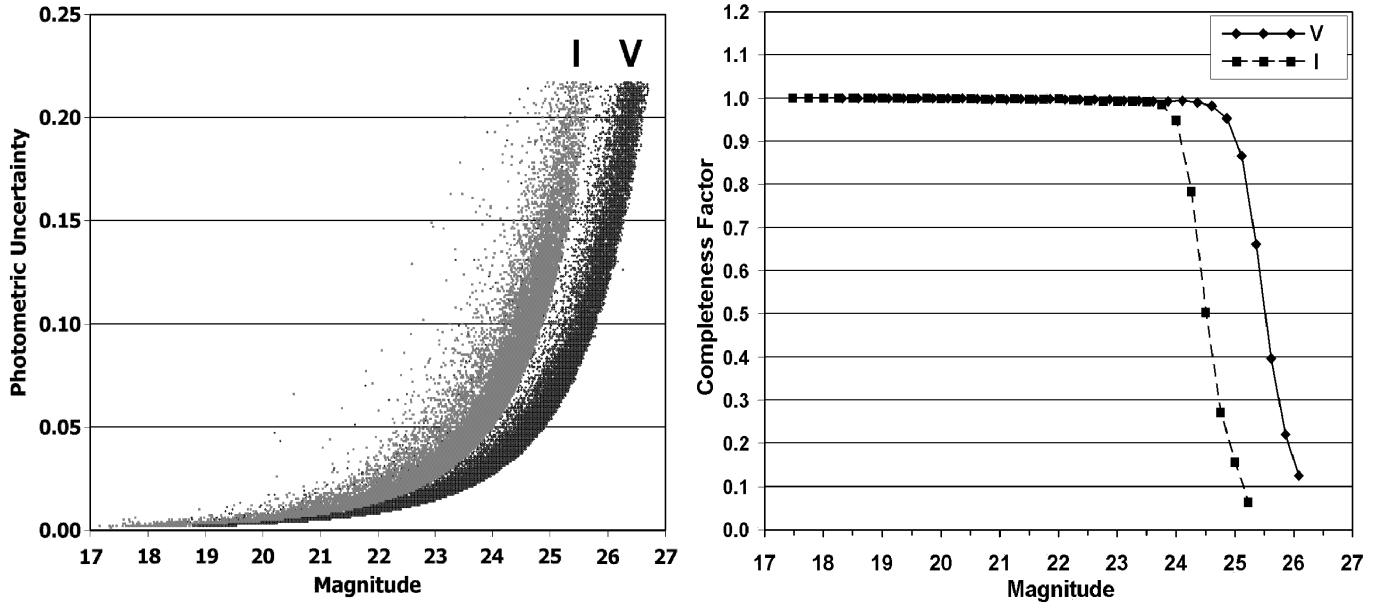

\centerline{\hbox{
\psfig{figure=f2a.ps,width=9.truecm,height=8.15truecm,angle=0}
\psfig{figure=f2b.ps,width=9.truecm,height=8.15truecm,angle=0}
}}
\caption{Uncertainties of photometry (left) and completeness factors (right) 
as derived by HSTphot from all six data sets, for both $F555W$ ($V$) and 
$F814W$ ($I$) bands.}
\label{fig_pherr-cmp}
\end{figure*}

Gouliermis et al. (2002) make a distinction between a known stellar
association in the LMC (LH 95), its surrounding field (located at the
periphery of the association) and the general background field of the
galaxy (farther away from the system). They found that the IMF of stars
with 3 \lsim\ $M/M{\solar}$ \lsim\ 10, in the area of the association has
a slope $\Gamma$ \simsim\ $-2$, while in the surrounding and remote fields
$\Gamma$ \simsim\ $-3$ and $-4$, respectively. According to Parker et al.
(2001) corrections for field star contamination of the IMFs of
associations can reduce an apparent slope $\Gamma$ from $-1.7$ to $-1.35$.
Hence, it should be kept in mind that inadequate corrections for
background stars may result in a steeper-than-in-reality IMF. Indeed
according to Gouliermis et al. (2002) the slope of the field-subtracted
IMF of the association LH 95 is found to be $\Gamma \simeq -1.6 \pm 0.3$
for 3 \lsim\ $M/M${\solar} \lsim\ 10, which becomes even more shallow with
$\Gamma \simeq -1.0 \pm 0.2$ if a wider mass range is taken into account
(3 \lsim\ $M/M${\solar} \lsim\ 28). For comparison, Garmany et al. (1982)
found $\Gamma=-2.1$ outside the solar circle and $\Gamma=-1.3$ inside, for
stars more massive than 20 M$_\odot$ within 2.5 kpc of the Sun. Later
Casassus et al. (2000) found little difference between the IMF slope
inside and outside the solar circle, being steeper than Salpeter's with
$\Gamma\sim-2$ everywhere, in line with previous results, according to
which the local field IMF is found to have a slope $\Gamma\sim-1.7$ to
$-1.8$ (Scalo 1986; Rana 1987; Kroupa, Tout, \& Gilmore 1993). Kroupa \&
Weidner (2003) suggest that superposition of Salpeter IMFs with a cluster
mass function slope of $-2.2$ explains the $\Gamma\sim-1.8$ field star
IMF.

Stellar cluster formation theories for turbulent molecular clouds show
that many processes are operating simultaneously, and it may be difficult
to find out which particular process dominates during the formation of a
particular star cluster. However, these theories consider partitioned
IMFs. For example, Bate et al. (2002) have distinguished brown dwarf
formation from that of other stars, and Gammie et al. (2003) have shown
that the high-mass part of the IMF gets shallower with time as a result of
coalescence and enhanced accretion. According to Elmegreen (2004) there is
an advantage in viewing the IMF in a multi-component way, because it
allows observers to anticipate and recognize slight variations in the IMF
for different classes of regions when they are sampled with enough stars
to give statistically significant counts. The same author discuss the
differences between the IMFs observed for clusters and remote fields and
suggests a three-component model of the IMF to consider possible origins
for the observed relative variations in brown dwarf, solar-to-intermediate
mass, and high-mass populations. He finds that these variations are the
result of dynamical effects that depend on environmental density and
velocity dispersion. Elmegreen's models accommodate observations ranging
from shallow IMF in cluster cores to Salpeter IMF in average clusters and
whole galaxies to steeper IMF in remote field regions.

In this study we present an example for the local variations of the slope
of the present-day mass function (PDMF or MF) in the general field of the
LMC, with high statistical significance, based on HST/WFPC2 observations
on six sequential fields located at the inner disk of the galaxy close to
the edge of its bar. The observed change of the MF slope is verified
statistically and it shows a trend of the MF to become flat for sub-solar
masses. Since these stars did not have the time to evolve, the MF in the
range $M$ \lsim\ 1 $M${\solar} accounts for the IMF of the LMC field.
We reconstruct the IMF of the field for higher masses after correcting for
evolutionary effects and we find that the change of the slope at about one
solar mass stands also for the LMC field IMF up to about 2.5 $M${\solar}.

The outline of this paper is the following: In the next section (\S 2) we
present our data and we describe the performed photometry. The star
formation history, as it was defined from previous studies is presented in
in \S 3, where the investigation of the observed populations takes place
and the observed color-magnitude diagram (CMD) is presented. The
determination and the study of the LF and the MF of the main-sequence
stars in the area is described in \S 4, where we also present the IMF of
the general LMC field of the inner disk, as it is reconstructed after
several assumptions have been taken into account.  General conclusions of
this investigation are presented in \S 5.

\section{Observations and Data Reduction}

The studied area is located in the inner disk of the galaxy, very close to
the western edge of the bar. The WFPC2 images of this area were collected
as target of the {\em Hubble Space Telescope} program GO-8576. Six
sequential telescope pointings were obtained, which cover a line from
North-East to South-West. In total the integration time is 2,000 seconds
observed in the {\em F555W} filter ($\sim V$) and 2,100 seconds in {\em
F814W} ($\sim I$) for each pointing. The exposure times are given in Table
1, together with other details of the observations. The WFPC2 fields,
overlayed on a SuperCOSMOS Sky Survey image of the general area, are shown
in Figure \ref{fig_map}. The photometry has been performed using the
package HSTphot as developed by Dolphin (2000a).  We added the individual
exposures in each filter for each field, with the use of the subroutine
{\em coadd}, to construct deep ones. These exposure times cover the high
dynamic range of brightness of the stars in the region with good overlap
between sets. We compared the long exposures with the shorter ones to
check for saturated stars and it seems that the brightest stars are within
the observed dynamic range. The detection limit of these observations is
at $V \sim$ 26.5 mag.

For the data reduction and the photometry we used the most recent version
of the HSTphot package (version 1.1.5b; May 2003). This version allows the
simultaneous photometry of a set of images obtained by multiple exposures.
HSTphot uses a self-consistent treatment of the charge transfer efficiency
and zero-point photometric calibrations and it is tailored to handle the
under-sampled nature of the point-spread function (PSF) in WFPC2 images
(Dolphin 2000b). We followed the standard procedures for removing bad
columns and pixels, charge traps and saturated pixels (subroutine {\em
mask}), and for the removal of cosmic rays (subroutine {\em crmask}) as
described by Dolphin (2000a). Since HSTphot allows the use of PSFs which
are computed directly to reproduce the shape details of star images as
obtained in the different regions of WFPC2, we adopted the PSF fitting
option in the photometry subroutine ({\em hstphot}), instead of performing
aperture photometry. Data quality parameters for each detected source 
are returned from {\em hstphot}, and we selected the values, which account 
for the best detected stars in uncrowded fields, as recommended in ``{\em 
HSTphot User's Guide}'' (object type 1 or 2, sharpness between $-0.3$ 
and $0.3$ and $\chi \le 2.5$).

HSTphot provides directly charge transfer efficiency corrections and
calibrations to the standard $VI$ system (Dolphin 2000b). Figure
\ref{fig_pherr-cmp} (left panel) shows typical uncertainties of photometry
as a function of the magnitude for the two filters. The completeness of
the data was evaluated by artificial star experiments, which were
performed for every observed field separately with the use of the HSTPhot
utility {\em hstfake}. The completeness was found not to change at all
from one frame to the other. The overall completeness functions are shown
in Figure \ref{fig_pherr-cmp} (right panel) for both filters. The chip of
WFPC2 consists of four frames, one of which (PC frame) has double
resolution but half field-of-view than the remaining three (WF frames).  
Our artificial-stars test showed that the completeness within the PC frame
of each camera pointing is somewhat lower than the one of the WF frames.
We noticed that this phenomenon is related to the small numbers of
detected stars in the PC frames, which are smaller than the ones of the
corresponding WF frames of the neighboring pointings, which overplot them.
Consequently, for this study we make use of the stars found only in the WF
frames of the six observed WFPC2 fields, which provide us with better
number statistics.


\begin{figure*}[t!]
\centerline{\hbox{
\psfig{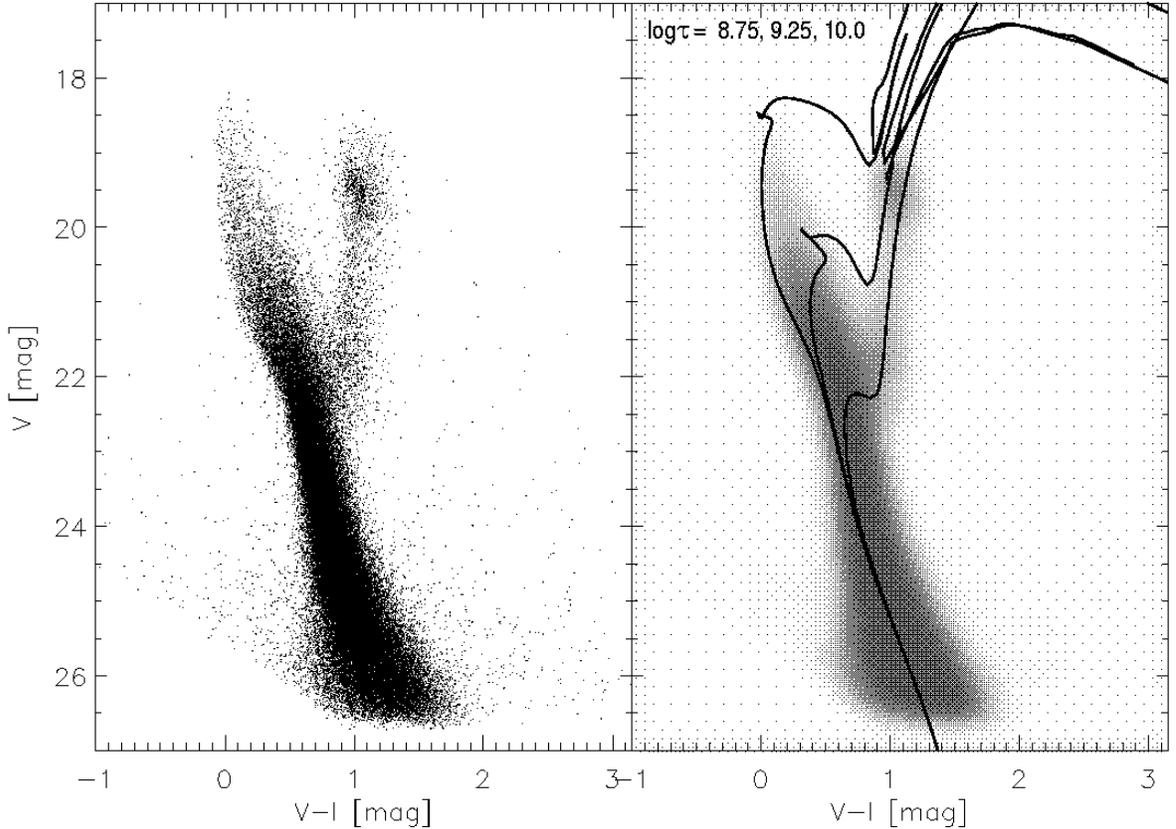}
}}
\caption{Left: $V-I$, $V$ Color-Magnitude Diagram (CMD) of the stars in 
all six observed WFPC2 fields. Right: Smoothed CMD of the same stars with 
reddened isochrone models from Padova group overplotted.}
\label{fig_cmdiso}
\end{figure*}

\section{Stellar Populations in the LMC Field}

The area investigated here accounts for the background field population of
the LMC inner disk close to the western edge of the bar of the galaxy. We
performed star counts on the stellar catalog of every one of the six
observed frames, to check for any stellar overdensity, which would be
expected in case of a concentration of stars, and would be an indication
of the existence of a stellar system. There are a few stellar clumps, of
which none shows a stellar density higher than 3$\sigma$ over the
background density (where $\sigma$ is the fluctuation in stellar density).
The observed area is checked with SIMBAD for known stellar systems and
only one is found, which is located at the south-eastern part of pointing
No. 6 (Figure \ref{fig_map}), and which has been identified as a star
cluster by Kontizas et al. (1990; KMHK 448), but it has not been revealed
with a stellar density higher than 3$\sigma$ in our observations.
Consequently there is no indication of the co-existence of different
stellar populations in the observed area, which thus can be considered to
be representative as a whole of the field population of the inner disk of
the LMC.

\subsection{Star Formation History of the LMC}

The general background field of the LMC has been previously observed with
HST/WFPC2 by several authors. These studies, having concentrated on the
star formation history (SFH) of the galaxy, give consistent results. The
first such investigation is by Gallagher et al. (1996), who observed a
field at the outer disk of the LMC, while a field in the inner disk of the
galaxy was observed by Elson et al. (1997). Three fields of the outer disk
were studied by Geha et al. (1998). Two of those fields are also presented
by Holtzman et al. (1999) with an additional field located in the LMC bar.
More recently, Castro et al. (2001) observed seven additional fields, all
located at the outer disk. Smecker-Hane et al. (2002) studied a field at
the center of the LMC bar and another one in the inner disk. The most
recent investigation, which makes use of such data is by Javiel et al.  
(2005), who use the same data material as Castro et al., to present an
analysis of the SFH of the LMC. These studies cover the whole sample of
investigations on the SFH of the galaxy with HST/WFPC2 observations, and
they cover a sample of 14 WFPC2 fields, which are spread in a large area
of the LMC, from the center of the bar to the edge of the outer disk.

The results of these investigators of the SFH in the disk (outer/inner)  
and the bar of the LMC will be summarized in the following: In general the
LMC field is dominated by an old stellar population with $\tau$ \gsim\
10 Gyr (Castro et al. 2001). A major increase in the star formation rate
(SFR)  occurred $\sim$ 2 Gyr ago, which resulted in almost 25\% of the
field stellar population, including much of the LMC disk (Gallagher et al. 
1996). Elson et al. (1997) found that an intense star formation event, 
which occurred $\sim$ 2 - 4 Gyr ago, probably corresponds to the formation 
of the disk. An increase (by a factor of 3) of the SFR, almost 2 Gyr ago, 
was also suggested by Geha et al. (1998). According to the same authors a
closed-box chemical evolution scenario implies that the LMC metallicity
has been doubled the past 2 Gyr. Events of enhanced SF are found to have
taken place in the north, north-west regions of the disk 2 - 4 Gyr ago
also by Castro et al. (2001). Javiel et al. (2005) using the same
observational material found slightly earlier events of enhanced SF at 1
to 2 or 3 Gyr ago for the north-western regions, located at the outer LMC
disk. The SFH of the LMC disk seems to have remained continuous with
almost constant SFR over the last $\sim$ 10 Gyr (Geha et al. 1998) or 15
Gyr (Smecker-Hane et al. 2002). The bar of the LMC has been found to
contain a larger number of older stars than in the disk (Holtzman et al.
1999) and according to Smecker-Hane et al.  (2002) it probably formed
during an episode that occurred 4 - 6 Gyr ago. These authors also suggest
that both the bar and disk of the LMC experienced similar SFHs at $\sim$ 
7.5 to 15 Gyr. It is worth noting that active SF during the past $\sim$ 
0.1 - 1 Gyr was also found in the disk by Gallagher et al. (1996). The
conclusive results presented here on the SFH of the galaxy will provide a 
baseline for the interpretation of the features observed in the
color-magnitude diagram of the area, presented in the next section.

\subsection{Color-Magnitude Diagrams}

Each of the observed WFPC2 fields includes around 13,500 stars. Hence the
overall stellar catalog, which includes almost 80,000 stars, provides a
rich sample of the LMC disk population. The $V-I$ vs. $V$ Color-Magnitude
diagram (CMD) of these stars is shown in Figure \ref{fig_cmdiso} (left
panel). As can be seen from this CMD the LMC field is characterized by a
prominent red-giant clump located around $V$ $\simeq$ 19.5 mag and $(V-I)
\simeq$ 1.1 mag, and a large number of low-mass MS stars below the
turn-off point of the $\sim$ 10 Gyr model. These features are also
apparent in the CMDs presented in the investigations of the SFH of the
LMC, mentioned above. Furthermore, there is a lack of main-sequence stars
brighter than $V \sim$ 18 mag in all CMDs. We verified that in our fields
this is not because of saturation. Indeed, the LMC field is known to have
only a few massive MS stars per unit area (Massey et al. 1995).  The
similarity of the CMDs of all these various WFPC2 fields suggests that the
stellar content of the general LMC field does not change significantly
from one area to the next, although the SFH of the bar has been found to
be somewhat different to the one of the disk of the galaxy, as we
discussed above. However, our fields are only in the disk and not in the 
bar.

\begin{figure*}[t!]
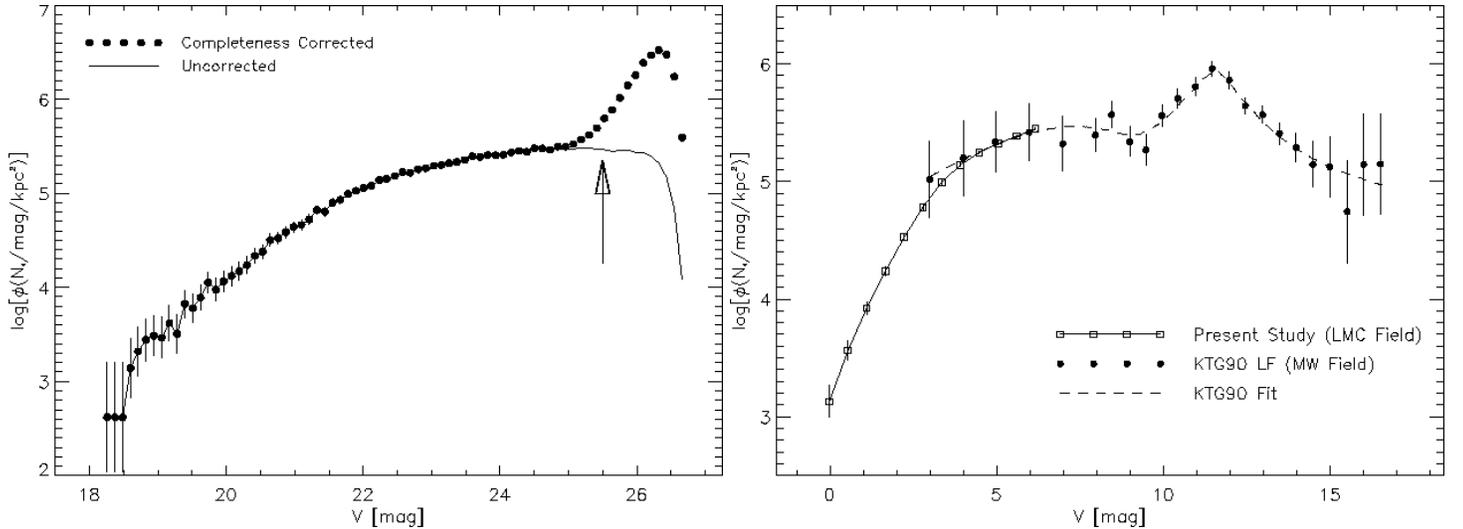

\centerline{\hbox{
\psfig{figure=f4a.ps,width=9.5truecm,angle=0}
\psfig{figure=f4b.ps,width=9.5truecm,angle=0}
}}
\caption{Left: Main-sequence Luminosity Function of the stars in all six 
observed fields. The arrow indicates the 50\% completeness level. Right: 
Comparison between the LF for the stars in the LMC field observed in this 
study within the 80\% completeness limit (open points) and the one for 
field stars near the Sun from Scalo (1989) and Stobie et al.  (1989) 
(filled points) together with the corresponding smooth fit (dashed line) 
adopted by Kroupa et al. (1990; KTG90). It is interesting to note that our 
LF for the low-mass stars is in very good agreement with the LF of the 
solar neighborhood in their region of overlap.} \label{fig_wflf}
\end{figure*}

There are also variations in the stellar density of the LMC field between
different areas. These variations are found to be due to a gradient in the
number of stars, with higher numbers towards the center of the galaxy
(Castro et al. 2001). Each of the areas observed by Smecker-Hane et al.
(2002)  in the bar and the inner disk of the galaxy (1\farcd7 southwest of
the center of the LMC) covers $\simeq 10^{5}$ stars, which correspond to
about 15,000 stars per WFPC2 field. These stellar numbers are in agreement
with the ones of the fields presented here and observed within the same
HST proposal ($\sim$ 13,500 stars per WFPC2 field) and the $\sim$ 15,800
stars observed by Elson et al. (1997) in another WFPC2 field at the inner
disk. On the other hand each of the WFPC2 fields at the outer disk studied
by Castro et al. (2001) contains around 2,000 stars in agreement with the
number of detected stars by Gallagher et al. (1996) in their single field,
which is also located in the outer disk. The number of stars in the
general LMC field found in one WFPC2 field earlier by us (Gouliermis et
al. 2005; $\sim$ 4,000 stars) is in line with these variations, since this
field is not close to the bar, but still much closer to it than all the
previously mentioned outer disk fields.

Concerning the SFH of the area presented here, the information provided in
the previous section is more than sufficient for the age distribution of
the contributing populations in the CMD of Figure \ref{fig_cmdiso} to be
accurately estimated.  In the same figure (right panel) we present a
smoothed image of the CMD with three indicative isochrone models
overplotted. The models of the Padova group in the HST/WFPC2 magnitude
system (Girardi et al. 2002) were used. These isochrone fits indeed show
that the stellar populations of the area seem to be the product of star
formation events that took place 0.5 - 3 Gyr ago. In addition, the
sub-giant region of the CMD is very well traced by models of older ages
(\gsim\ 10 Gyr).  It should be noted that according to the tabulation
by Ratnatunga \& Bahcall (1985), the contamination of the CMD by Galactic
foreground stars is expected to be negligible. Furthermore, the
investigation of Metcalfe et al. (2001)  on the Hubble Deep Fields, has
shown that also the number of background faint galaxies in such a CMD, is
very small for stars brighter than V $\simeq$ 25 mag.

Isochrone fitting showed that the color excess toward every one of the six
observed fields is almost the same and it can be considered as uniform for
the whole area. The mean value of the color excess from isochrone fitting
was found to be $E(V-I)\simeq$ 0.05, which corresponds to $E(B-V)\simeq$
0.03 (e.g. Rieke \& Lebofsky 1985). The reddening curve $R_{V}$ (ratio of
the total absorption in $V$, $A_{V}$, to $E(B-V)$), has been found to be
consistent by various authors (Mihalas \& Binney 1981: $R_{V}=3.2$;
Koornneef 1983: $R_{V}=3.1$; Leitherer \& Wolf 1984: $R_{V} \simeq 3.13$).
Adopting the value estimated by Taylor (1986; $R_{V}=3.15$), the
absorption toward this area is $A_{V} \simeq 0.095$ mag. The models in
Figure \ref{fig_cmdiso} are reddened accordingly. We adopted the distance
modulus for the LMC derived by Panagia et al. (1991) from SN 1987A, 18.5
$\pm$ 0.1 mag, which corresponds to a distance of 50.1 $\pm$ 3.1 kpc.

\section{Luminosity and Mass Function of the Field}

\subsection{Luminosity Function}

We examine the main-sequence luminosity function (LF) of the stars found
in all six observed fields (overall catalog). We constructed the LF of the
stars based on the WF frames of the WFPC2 fields only, due to their better
completeness, as mentioned earlier. The main-sequence stars were selected
according to their positions in the CMD and their LF, ${\Phi}$, was
constructed by counting them in magnitude bins and normalizing their
numbers to the same surface of 1 kpc$^{2}$. The constructed LF is shown in
Figure \ref{fig_wflf} (left panel). The solid line represents the LF
uncorrected for completeness. The completeness corrected LF is plotted
with thick dots. As shown by the completeness functions of Figure
\ref{fig_pherr-cmp}, the stellar sample down to $V \simeq 25$ mag is
complete by 95\% (which corresponds to $I \simeq 24$ mag). Then the
completeness falls rapidly to almost 50\% within 0.5 magnitudes. The steep
decline of the completeness for stars fainter than $V \simeq$ 25 mag
introduces the sharp rise of the completeness-corrected LF toward the
faint end. Since only data within the 50\% completeness limit can be used
this rise affects only the last three useful magnitude bins (the arrow in
Figure \ref{fig_wflf} indicates the 50\% completeness limit).

The comparison of this LF with the ones presented by Smecker-Hane et al.
(2002) of their ``Disk 1'' field and the bar region shows that these LFs
are similar to each other in various aspects. Specifically, in the LF of
Figure \ref{fig_wflf} one can observe a change of its slope at $V \sim$
22.2 - 22.4 mag, like in both bar and disk LFs of Smecker-Hane et al.
(2002). This magnitude corresponds to the turn-off for stars of age $\sim$
2 Gyr and this change is much smoother than in the case of Smecker-Hane
et al. LFs. In addition, we observe changes in the slope of our LF, which
seem to coincide with two of the three ``spikes'' (as Smecker-Hane et al.
call them) observed in the bar LF at almost the same magnitudes ($V
\simeq$ 19.6 and 20.4 mag).  This shows that the LF of our area, which
belongs to the inner disk, is more similar to the one of the bar than of
the outer disk. This is more or less expected, considering that our area
is close to the edge of the bar (0\farcd3 away), while the ``Disk-1''
field of Smecker-Hane et al. (2002)  is located almost at the edge of the
inner disk, about 1\farcd5 away from the bar.

It is worthwhile to compare our LF with the one of low-mass field stars in
the solar neighborhood, as it is constructed with data taken from Stobie
et al. (1989) (for $M_{V} > 7$ mag) and Scalo (1989) ($7 \geq M_{V}/{\rm
mag} \geq 3$). To do so we resample our data into wider bins, and we
construct the LF of our LMC field with a sampling similar to the one of
the solar neighborhood LF, as it is presented by Kroupa et al. (1990).
This comparison, which is shown in Figure \ref{fig_wflf} (right panel)
shows that the LF of the low-luminosity stars seems to be the same between
the LMC general field and the local field of the Milky Way for 2.5
\lsim\ $M_{V}/{\rm mag}$ \lsim\ 6.5. This magnitude range covers the
brighter part of the flattening of the local neighborhood LF, which
corresponds to absolute magnitude $M_{V} \sim + 7$ mag and which is the
result of effects of H$^{-}$ molecules on the opacity and equation of
state (Kroupa et al. 1990). These effects together with the ones of
H$_{2}$ and other molecules introduce points of inflexion in the
mass-luminosity relation, at which the dominant opacity source in the
stellar atmosphere changes rapidly. The solar neighborhood LF of Figure
\ref{fig_wflf} has a maximum near $M_{V} \sim 12$ mag, which according to
Kroupa et al. (1990) is consistent with being a consequence of a point of
inflexion in the mass-luminosity relation, caused by the effect of H$_{2}$
on stellar photosphere. Our data do not allow us to compare these LFs
toward fainter magnitudes.

\begin{figure*}[t!]
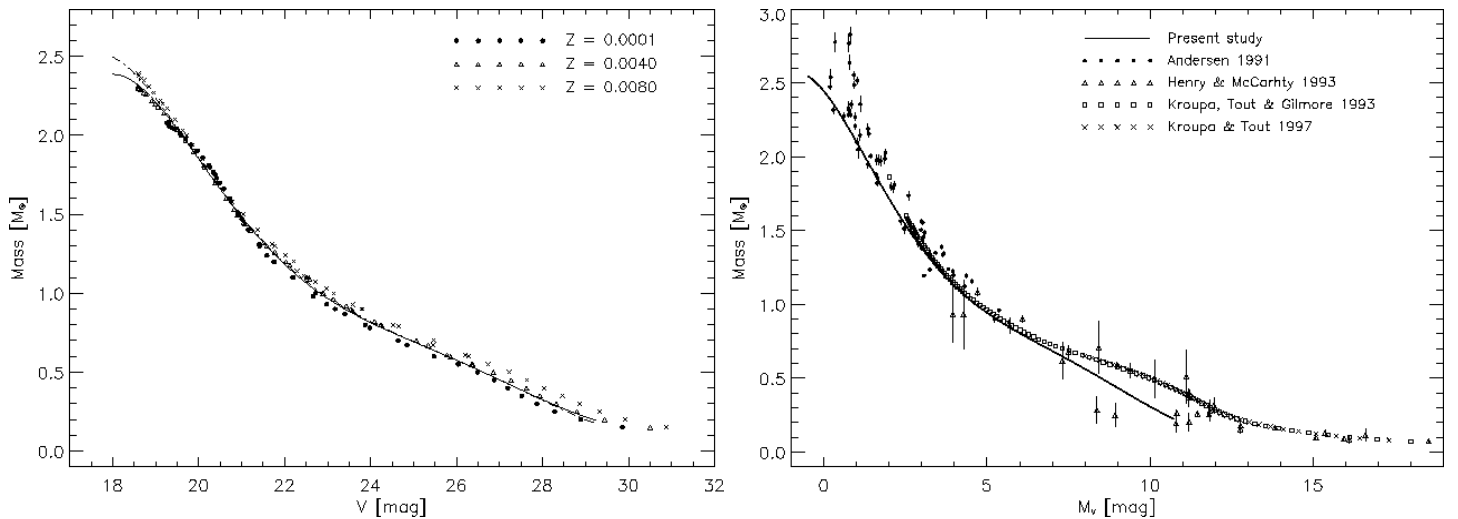

\centerline{\hbox{
\psfig{figure=f5a.ps,width=9.5truecm,angle=0}
\psfig{figure=f5b.ps,width=9.5truecm,angle=0}
}}
\caption{Left: Mass-to-Luminosity Relation, constructed in this study
according to three isochrone models and for three assumed metallicities.  
Right: Comparison of this relation, adopted here, with previous empirical
and semi-empirical relations.} \label{fig_m2lr}
\end{figure*}

\subsection{Mass Function}
\subsubsection{Fundamental Definitions}

The distribution of stellar masses calculated for a given volume of space
in a stellar system is known as the Present Day Mass Function (PDMF),
which we usually call the Mass Function of the system. In the case of the
area presented here, there is no specific stellar system to which the
observed stars seem to belong and the distribution of their masses
represents the PDMF of the field population of the inner LMC disk. We will
refer to the function $\xi(\log{m})$, which gives the number of stars per
unit logarithmic (base ten) mass interval $d\log{m}$ per unit area (e.g.
/kpc$^{2}$) as the {\em mass function} (MF) of the field. This function
usually replaces the {\em mass spectrum} $f(m)$, which is the number of
stars per unit mass interval $dm$ per unit area. The characterization of
these functions is based on the various parameterizations used for the
Initial Mass Function (IMF) of a stellar system (see e.g. Kroupa 2002).
Very useful parameters are the indices of the mass spectrum $f(m)$, and of
the mass function $\xi(\log{m})$, defined as \beq \gamma =
\frac{d\log{f(m)}}{d\log{m}} \eeq \label{eq-fga} and \beq \Gamma =
\frac{d\log{\xi(\log{m})}}{d\log{m}} \label{eq-xiga}\eeq These are the
logarithmic slopes of $f(m)$ and $\xi(\log{m})$ and for power-law mass
spectra they are independent of mass (Scalo 1986). A reference slope is
the logarithmic derivative $\Gamma = -1.35$, which is the index of the
classical IMF for stars in the solar neighborhood with masses 0.4
\lsim\ $M/M${\solar} \lsim\ 10, found by Salpeter (1955). The
corresponding mass spectrum has $\gamma = \Gamma - 1 \simeq -2.35$. The
lognormal field star IMF fit by Miller \& Scalo (1979) has $\Gamma \simeq
-(1+\log{m})$. A basic relation between $\xi(\log{m})$ and $f(m)$ is
$\xi(\log{m}) = (\ln{10})\cdot m \cdot f(m) \simeq 2.3 \cdot mf(m)$ (see
Scalo 1986).

\subsubsection{Construction of the Mass Function}

In the present study we construct the MF $\xi(\log{m})$ by counting stars
in logarithmic (base ten) mass intervals. The counting of stars in mass
intervals can be achieved by translating their luminosities (or
magnitudes) into masses using mass-luminosity relations and then
constructing the distribution of the derived masses. This method is
definitely dependent on the adopted evolutionary models used for the
transformation of luminosities to masses. de Grijs et al. (2002b), who
presented a thorough investigation of the Mass-Luminosity (M/L) relation
found that this model dependence results only in a systematic offset of
the overall masses.

For the determination of the M/L relation we used the latest Padova
theoretical isochrones in the HST/WFPC2 Vega system (Girardi et al. 2002).  
These models were originally developed by Girardi et al. (2000) and they
were transformed to the HST/WFPC2 pass-bands as described in Salasnich et
al. (2000). The M/L relation of our study here is defined by the models of
three different ages. These models (overplotted over the CMD in Figure
\ref{fig_cmdiso} - right panel) are carefully selected as the most
representative of the age limits estimated for the observed field
population.  These limits, which have been verified by previous
investigations of the SFH of the general LMC field population (see \S
3.1), cover an age span between $\sim$ 0.6 - 10 Gyr (8.75 \lsim\
$\log{\tau}$ \lsim\ 10). The M/L relation for the main-sequence stars
in the observed CMD was developed according to the main-sequence part of
the evolutionary models. For the low main-sequence population the 10 Gyr
model provided the M/L relation for stars up to $M_{\rm V} \simeq 6.95$
mag (0.7 $M${\solar}). A model, which corresponds to a median age ($\sim$
1.8 Gyr) was used for the M/L relation for stars with magnitudes from
$M_{\rm V} \simeq 6.97$ mag (0.7 $M${\solar} according to this model) up
to $M_{\rm V} \simeq 3.8$ mag ($\sim$ 1.1 $M${\solar}).

\begin{figure*}[t!]
\centerline{\hbox{
\psfig{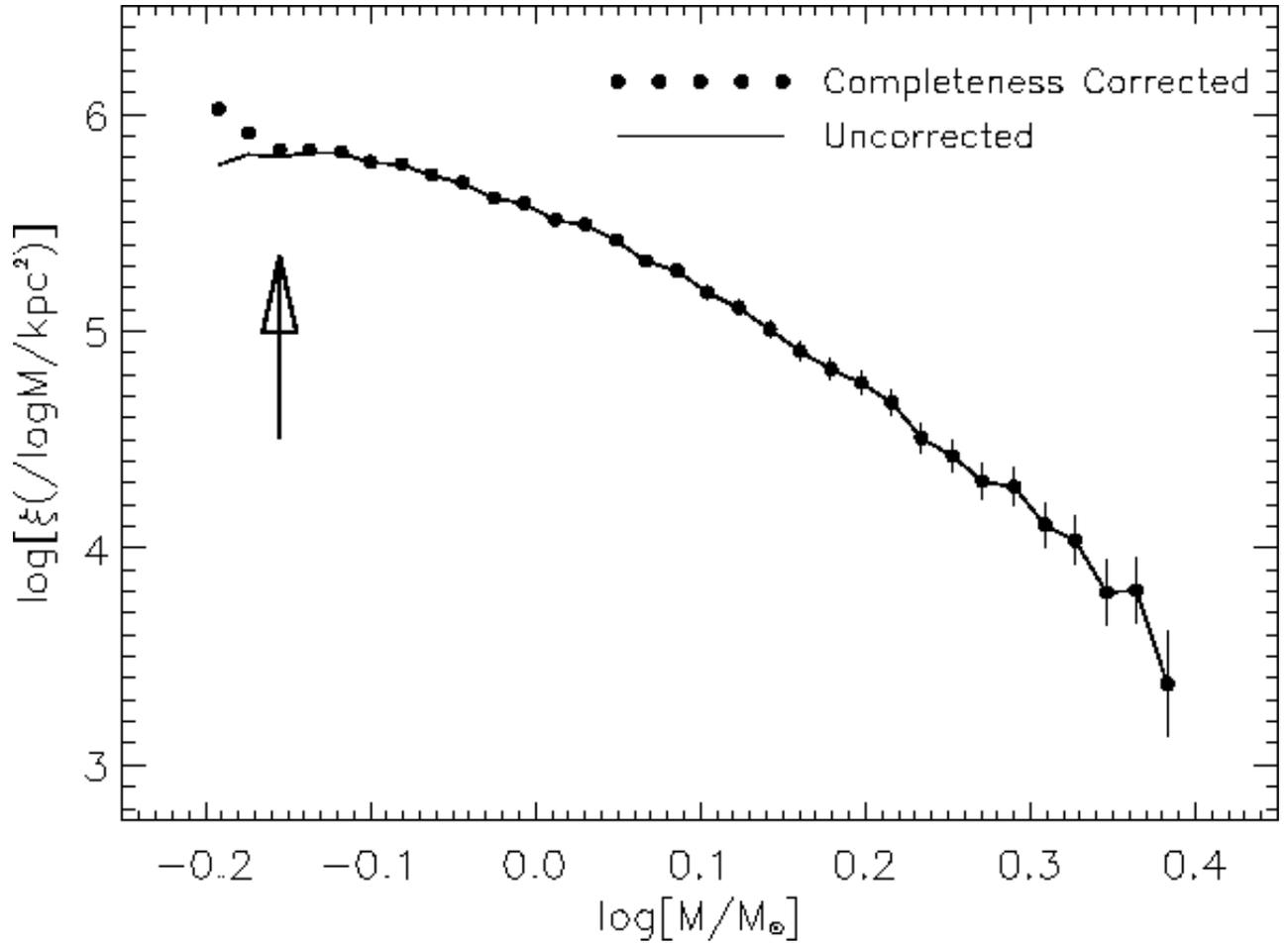}
}}
\caption{Main-sequence Mass Function of the stars detected in the WF
frames of all six observed fields, for completeness \gsim\ 50\%. The
arrow indicates the $\sim$ 95\% completeness limit.} \label{fig_wfmf}
\end{figure*}

For the upper main-sequence stars in the observed fields, which correspond
to the younger population, the 0.6 Gyr model was used. The use of any
younger isochrone would have no meaning, because there are no stars in the
CMD of Figure \ref{fig_cmdiso} brighter than the turn-off of the 0.6 Gyr
model, also considering that there are no saturated stars in any observed
field. Therefore, the highest mass that could be measured from these data
according to the $\sim$ 0.6 Gyr isochrone is \lsim\ 2.5 $M${\solar} (which
corresponds to $M_{\rm V} \simeq 0.06$ mag). In all models the LMC
metallicity of Z $=$ 0.008 was taken into account. This value is in the
range of values 0.002 $<$ Z $<$ 0.018 compiled by Kontizas et al. (1993)  
from the literature. In order to check for any biases in the determination
of the M/L relation due to an adopted wrong metallicity, the procedure
described above was repeated also for models of lower metallicities
(Z=0.004 and Z=0.0001). We found that there are {\em no significant
differences between the derived M/L relations from models of different
metallicities}. This is shown in Figure \ref{fig_m2lr} (left panel), where
the relation of magnitudes and masses for each model has been plotted.
Three different symbols represent the M/L relations derived from models of
different metallicity. The corresponding polynomial fits on the data for
Z=0.008 (dashed line) and on all the data (solid line) are overplotted.
These fits differ slightly only at the higher mass by 0.1 $M${\solar}. The
overall polynomial fit (solid line) is the adopted M/L relation for the
construction of the MF of the LMC field in this study.

This fit is plotted in Figure \ref{fig_m2lr} (right panel), along with M/L
relations derived from previous empirical (Andersen 1991; Henry \&
McCarthy 1993) and semi-empirical (Kroupa et al. 1993; Kroupa \& Tout
1997) studies for solar-metallicity stellar populations for $M$ \lsim\
2 $M${\solar}. For solar metallicities in the mass range 0.1 $<$
$M/M{\solar}$ \lsim\ 1 Kroupa \& Tout (1997) conclude that the
theoretical M/L relations of Baraffe et al. (1995) provide the best
overall agreement with all recent observational constraints. Their
semi-empirical M/L relation though (plotted in the right panel of Figure
\ref{fig_m2lr}), is closely followed by the theoretical M/L solar
abundance relation by Chabrier et al. (1996) (see also de Grijs et al.  
2002b). The best match with the observational data for low-metallicity
Galactic globular clusters for masses $M$ \lsim\ 0.6 - 0.8 $M{\solar}$
is found by Piotto et al. (1997) to be provided by the theoretical M/L
relations of Alexander et al. (1997).  The comparison of our M/L relation
based on the Padova models with the previous studies shows a very good
agreement for the magnitude range of interest ($M_{\rm V}$ \lsim\ 6.5
mag), although there is a small difference toward the bright end of the
plot (for 0 \lsim\ $M_{\rm V}$/mag \lsim 2.5). In addition there is a
slight offset of our fit toward smaller masses.

Here we have a final comment on the constructed M/L relation and the
resulting MF. We checked both the present-day and initial stellar masses
of each magnitude as they are provided by the models and we found that
they are exactly the same for stars with magnitudes within the limits
covered by the part of the main-sequence of each model used for the
construction of our M/L relation.  This fact implies that evolutionary
effects did not reduce the stellar masses through mass loss. Consequently
all used mass bins for the construction of the MF correspond to both the
present-day and initial stellar masses. This will be useful later for the
reconstruction of the IMF of these stars.

\subsubsection{Linear Regression}

The constructed MF of the main-sequence stars in the observed fields is
shown in Figure \ref{fig_wfmf}. The stars were counted in logarithmic mass
intervals according to their masses as they were estimated with the use of
the M/L relation presented in the previous section.  The stellar numbers
are normalized to a surface of 1 kpc$^2$. In Figure \ref{fig_wfmf} the MF
not corrected for incompleteness is represented by a line, and points
indicate the corrected MF. Only data within 50\% of completeness are
plotted and the arrow indicates the $\sim$ 95\% completeness limit.

The very fine binning in this MF gives the impression that this
distribution cannot be considered as a unique power-law throughout its
whole mass range. Indeed, it seems that there is a gradual change in the
slope of the MF similar to the one observed in the LF (see \S 4.1). In
order to check for variations in the slope of the MF for stars in
different mass ranges we modeled the MF data with a weighted linear fit
({\em linear regression}). Specifically, we considered the problem of
fitting a set of $N$ data points ($\log{m_{i}}$, $\log{\xi(\log{m_{i}})}$)
to a straight-line model \beq \log{\xi(\log{m})} = \Gamma \cdot \log{m} +
\beta \label{eq-fit}\eeq considering different ranges of mass bins
($\log{m_{i}}$) from the MF shown in Figure \ref{fig_wfmf}. For simplicity
we call the function $F(m) = \xi(\log{m})$. Hence Equation (\ref{eq-fit})  
becomes \beq \log{F(m)} = \Gamma \cdot \log{m} + \beta
\label{eq-fit2}.\eeq

The fit was applied for the data up to the highest available mass bin of
$M \simeq$ 2.4 $M${\solar}, but each time changing the considered low-mass
bin, down to the one, which corresponds to 50\% completeness ($M \simeq$
0.63 $M${\solar}). For every considered mass range we evaluated the MF
slope $\Gamma$ and we measured how well the model agrees with the data
using the chi-square merit function (Press et al. 1992), which in this
case is \beq \chi^{2}(\beta,\Gamma) = \sum^{N}_{i=1} \left(
\frac{\log{F_{i}}-\beta-\Gamma \cdot \log{m_{i}}}{\sigma_{i}} \right)^{2}
\label{eq-chi2}\eeq where $\sigma_{i}$ is the uncertainty associated with
each measurement $F_{i}$. We are interested to find the mass range within
which the linear model of Equation (\ref{eq-fit2}) is in the best
agreement with the data. Hence we estimated for every mass range the
goodness-of-fit of the data to the model, which is the probability $Q$
that a value of $\chi^{2}$ as poor as the value in Equation
(\ref{eq-chi2}) should occur by chance, given by the equation: \beq Q =
\mathcal{G} \left( \frac{N-2}{2},\frac{\chi^{2}}{2}\right)  
\label{eq-gof}\eeq where $\mathcal{G}$ is the incomplete gamma function
$Q(\beta,\log{m})$. The goodness-of-fit is believable if $Q$ is larger
than about 0.1, while for 0.001 \lsim\ $Q$ \lsim\ 0.1, the fit may
be acceptable if the errors are non normal (Press et al. 1992).

\begin{table*}[t]
\begin{center}
\caption{Slopes $\Gamma$ of the MF and the corresponding goodness-of-fit 
($Q$) for the linear regression (least $\chi^{2}$ method) for different 
mass intervals. With the application of linear regression for different 
mass ranges we found that the MF slope shows a statistically significant 
change at about 1 $M${\solar}. The rows marked with boldface characters 
indicate the mass range at which the fit starts to give believable results 
($Q$ \gsim\ 0.1). \label{tab2}}
\begin{tabular*}{\columnwidth}[]{@{\extracolsep{\fill}}ccccc}
\tableline
Mass Range & $N$ & $\Gamma$ & $\beta$ & $Q$\\
($M${\solar}) & & & &  \\
\tableline
\tableline
  0.64 -   2.42&  32&$-$2.91 $\pm$ 0.04& 9.04 $\pm$ 0.00& 0.000 \\
  0.67 -   2.42&  31&$-$2.94 $\pm$ 0.04& 9.04 $\pm$ 0.00& 0.000 \\
$\cdot$&$\cdot$&$\cdot$&$\cdot$&$\cdot$\\
$\cdot$&$\cdot$&$\cdot$&$\cdot$&$\cdot$\\
$\cdot$&$\cdot$&$\cdot$&$\cdot$&$\cdot$\\
  0.94 -   2.42&  23&$-$4.23 $\pm$ 0.09& 9.14 $\pm$ 0.01& 0.000 \\
  0.98 -   2.42&  22&$-$4.43 $\pm$ 0.10& 9.16 $\pm$ 0.01& 0.011 \\
{\bf 1.03 -   2.42}&{\bf   21}&{\bf $-$4.59 $\pm$ 0.11}&{\bf 9.19 $\pm$ 0.01}&{\bf  0.087} \\
  1.07 -   2.42&  20&$-$4.79 $\pm$ 0.12& 9.22 $\pm$ 0.02& 0.678 \\
  1.12 -   2.42&  19&$-$4.90 $\pm$ 0.14& 9.24 $\pm$ 0.02& 0.781 \\
  1.17 -   2.42&  18&$-$4.99 $\pm$ 0.16& 9.26 $\pm$ 0.02& 0.816 \\
  1.22 -   2.42&  17&$-$5.17 $\pm$ 0.18& 9.29 $\pm$ 0.03& 0.963 \\
  1.27 -   2.42&  16&$-$5.23 $\pm$ 0.21& 9.31 $\pm$ 0.04& 0.953 \\
$\cdot$&$\cdot$&$\cdot$&$\cdot$&$\cdot$\\
$\cdot$&$\cdot$&$\cdot$&$\cdot$&$\cdot$\\
$\cdot$&$\cdot$&$\cdot$&$\cdot$&$\cdot$\\
\tableline
\tableline
  0.64 -   1.03&  12&$-$2.13 $\pm$ 0.07& 9.13 $\pm$ 0.01& 0.000 \\
  0.67 -   1.03&  11&$-$1.88 $\pm$ 0.08& 9.14 $\pm$ 0.01& 0.000 \\
  0.70 -   1.03&  10&$-$1.87 $\pm$ 0.10& 9.14 $\pm$ 0.01& 0.003 \\
{\bf 0.73 -   1.03}&{\bf 9}&{\bf $-$2.08 $\pm$ 0.12}&{\bf 9.13 $\pm$ 
0.01}&{\bf  0.117} \\
  0.76 -   1.03&   8&$-$2.27 $\pm$ 0.14& 9.12 $\pm$ 0.01& 0.390 \\
  0.79 -   1.03&   7&$-$2.36 $\pm$ 0.18& 9.12 $\pm$ 0.01& 0.356 \\
  0.83 -   1.03&   6&$-$2.64 $\pm$ 0.23& 9.11 $\pm$ 0.01& 0.782 \\
$\cdot$&$\cdot$&$\cdot$&$\cdot$&$\cdot$\\
$\cdot$&$\cdot$&$\cdot$&$\cdot$&$\cdot$\\
$\cdot$&$\cdot$&$\cdot$&$\cdot$&$\cdot$\\
\tableline
\tableline
\end{tabular*}
\tablenotetext{}{{\sc Note:} The mass of $\sim$ 0.64 $M${\solar}
represents the 50\% completeness limit, and the 95\% completeness limit
corresponds to $\sim$ 0.70 $M${\solar}.}
\end{center}
\end{table*}

\subsubsection{Results for the Mass Function}

With this procedure we found that, for data within the 50\% completeness,
not the whole observed mass range shows a significant correlation between
the mass and the corresponding stellar numbers (both in logarithmic
scale). In Table 2, the results of the linear regression for selected
samples are shown. It is found that an acceptable linear correlation
between $\log{F(m)}$ and $\log{m}$ ($Q$ \gsim\ 0.1) starts to appear
for stars within mass ranges with a lower limit $M \simeq$ 1.0
$M${\solar}.  From this limit on and for mass ranges with lower limits of
higher masses the goodness-of-fit becomes continuously higher, showing a
better linear correlation. The highest mass limit used is $M \simeq 2.4$
$M${\solar}.

The next step is to repeat this procedure in order to define the mass
range with the best linear correlation for stars with masses smaller than
the limit of $\sim$ 1.0 $M${\solar}.  We found that the data are very well
linearly correlated for masses in the range starting at $\simeq$ 0.73
$M${\solar} and higher (second set of slopes in Table 2). The MF slope in
this mass range (0.73 \lsim\ $M/M_{\solar}$ \lsim\ 1.03) is $\Gamma
\simeq -2.1 \pm 0.1$ (steeper than Salpeter's IMF), in contrast to the
slope for masses 1.0 \lsim\ $M/M_{\solar}$ \lsim\ 2.4, which is
found to be steeper with $\Gamma \simeq -4.6 \pm 0.1$. We used as lower
mass limit 0.7 $M${\solar}, which corresponds to the completeness limit of
95\%, to perform an additional test. It should be noted that for smaller
masses the completeness falls rapidly (see \S 2), so that the completeness
corrections introduce a very steep distribution of the very low-mass data
(as shown in the LF of Figure \ref{fig_wflf}), which does not have
necessarily any physical meaning. Consequently, we do not consider the MF
with $M < 0.7$ $M${\solar} test.

Hence, considering the fixed lower-mass limit to be 0.7 $M${\solar}, we
performed the same procedure for smaller mass intervals and found that
intervals, which correspond to continuously smaller masses have a MF
slope, which changes gradually to become almost Salpeter's value ($\Gamma
\simeq -1.4$ for 0.7 \lsim\ $M/M_{\solar}$ \lsim\ 0.9) or even more
shallow. The estimated slopes are given in Table 3, which demonstrates
that the MF toward the smaller observed masses becomes flat.
Unfortunately, this result for the shorter mass ranges is based on small
numbers of bins, and thus on poor number statistics (last rows in Table
3). A turn-over of the MF cannot be observed, unless deeper data with
better completeness are used. This result, which is a clear indication of
a low-mass flattening of the MF, also indicates that the estimated MF
slope is very sensitive to the selected mass range.


\begin{table*}[t]
\begin{center}
\caption{Slopes $\Gamma$ of the main-sequence MF for the linear fits for 
different mass ranges, with lower mass limit at 0.7 $M${\solar}. It is 
shown that the MF slope is becoming gradually more shallow for smaller  
mass intervals toward the low-mass observable limit within the 95\% 
completeness. The boldface row indicates again the mass range at which the 
fit starts to be believable. \label{tab3}}
\begin{tabular*}{\columnwidth}[]{@{\extracolsep{\fill}}ccccc}
\tableline
Mass Range & $N$ & $\Gamma$ & $\beta$ &$Q$\\
($M${\solar}) & & & &\\
\tableline
\tableline
$\cdot$&$\cdot$&$\cdot$&$\cdot$&$\cdot$\\
$\cdot$&$\cdot$&$\cdot$&$\cdot$&$\cdot$\\
$\cdot$&$\cdot$&$\cdot$&$\cdot$&$\cdot$\\
  0.70 -   1.12&  12&$-$2.03 $\pm$ 0.08& 9.13 $\pm$ 0.01& 0.00 \\
  0.70 -   1.07&  11&$-$1.94 $\pm$ 0.09& 9.14 $\pm$ 0.01& 0.00 \\
  0.70 -   1.03&  10&$-$1.87 $\pm$ 0.10& 9.14 $\pm$ 0.01& 0.00 \\
  0.70 -   0.98&   9&$-$1.72 $\pm$ 0.11& 9.16 $\pm$ 0.01& 0.02 \\
  0.70 -   0.94&   8&$-$1.63 $\pm$ 0.13& 9.17 $\pm$ 0.01& 0.03 \\
{\bf 0.70 -   0.90}&{\bf 7}&{\bf $-$1.40 $\pm$ 0.16}&{\bf 9.20 $\pm$ 0.02}&{\bf  0.23}\\
  0.70 -   0.86&   6&$-$1.25 $\pm$ 0.19& 9.22 $\pm$ 0.02& 0.27 \\
  0.70 -   0.83&   5&$-$1.01 $\pm$ 0.25& 9.25 $\pm$ 0.03& 0.41 \\
  0.70 -   0.79&   4&$-$0.94 $\pm$ 0.35& 9.26 $\pm$ 0.05& 0.25 \\
  0.70 -   0.76&   3&$-$0.27 $\pm$ 0.54& 9.35 $\pm$ 0.07& 0.78 \\
\tableline
\tableline
\end{tabular*}
\tablenotetext{}{{\sc Note:} The mass of $\sim$ 0.70 $M${\solar} 
represents the 95\% completeness limit.}
\end{center}
\end{table*}

Taking into account that stars of small masses evolve very slowly one may
assume that the low-mass MF has a slope close to Salpeter's because it
actually accounts for the Initial Mass Function (IMF) of these stars, and
that the steepening of the higher mass MF is due to evolutionary effects.
We investigate this aspect in the next section, where we also attempt a
reconstruction of the IMF of the observed LMC field in a qualitative
manner, taking several assumptions into account.

\section{Reconstruction of the Initial Mass Function}

In this section we use the information on the (present-day) MF of the LMC
field in the observed area, to reconstruct its IMF. In order to achieve
this we have to make a number of assumptions concerning the evolution
process of the stars found in the area. An issue to be clarified first is
{\em if the present-day mass, as it was estimated from the models for each
main-sequence star, differs from the corresponding initial mass}.  
Considering that for stars in the observed mass range no mass loss takes
place we checked the models and we verified that indeed for each magnitude
bin used for the construction of the adopted M/L relation, {\em the
present-day stellar masses are exactly equal to their initial values for
the main-sequence stars}. This was verified for the models of all three
considered metallicities and for the mass ranges, for which each one of
the three adopted ages was used (0.2 \lsim\ $M/M${\solar} \lsim\ 0.7
for $\tau \simeq$ 10 Gyr, 0.5 \lsim\ $M/M${\solar} \lsim\ 1 for
$\tau \simeq$ 2 Gyr and 1 \lsim\ $M/M${\solar} \lsim\ 2.4 for $\tau
\simeq$ 560 Myr). 

Consequently, the masses estimated for the observed main-sequence stars,
which were used for the construction of their PDMF are actually the
initial masses of the stars (without taking the red giants into account)
and thus, they correspond also to their IMF. If mass-loss would occur, as
is the case for high-mass stars, then the present-day masses of the stars
would not be useful for the construction of their IMF, because the stars
would be distributed in mass bins different than those of their original
masses. Since this is not the case, the expected differences between the
PDMF and the IMF of the observed main-sequence stars are mostly due to
different stellar numbers and not due to different mass estimates, which
would redistribute the same stars in different mass bins. It should be
noted that the observed MF of the lower main-sequence stars, below a
specific turn-off, which did not have the time to evolve, should account
for the IMF of these stars. We discuss this in the following section.

\subsection{Assumptions concerning the small masses}

The first assumption that has to be made for the reconstruction of the IMF
considers the upper mass limit of the {\em lower main-sequence} stars.
Specifically, is should be defined which stars actually belong to the
``lower main-sequence''. A reasonable assumption is to define this upper
mass limit as the turn-off (MSTO) of the 10 Gyr isochrone. The choice of
this isochrone is based on the fact that this model seems to represent the
majority of the sub-giant branch shown in Figure \ref{fig_cmdiso}. Below
this magnitude there should be also primordial faint main-sequence stars,
but an examination of both the MSTO of the 10 Gyr model and the one of the
oldest available of 15.5 Gyr showed that they do not differ significantly.  
Specifically the two MSTOs differ by almost 0.4 magnitudes, the brighter
point being the value for the younger isochrone of course, which accounts
for 0.1 $M${\solar}. The (initial) mass of the MSTO for the 15.5 Gyr model
is around 0.85 $M${\solar}, and for the 10 Gyr is about 0.95 $M${\solar}.
Hence, we can safely state that {\em the observed PDMF accounts for the
IMF of stars with masses up to $\simeq$ 0.9 $M${\solar}}.

It should be noted that while the exact details of the star formation
history (SFH) of the LMC are still under debate, various authors suggest
that the \gsim\ 12 Gyr old population does not contribute more than
$\approx$ 5\% to the total low-mass stellar content of the LMC, and that
the past 2 to $\approx$ 7 Gyr saw several epochs with strongly enhanced
star formation rates (see, e.g, Gallagher et al.\ 1996; Elson et al.\
1997; Geha et al.\ 1998). A 5\% contribution of \gsim\ 12 Gyr old stars to
the stellar population should only have a minor effect on the overall
slope of the present day mass function. Hence we suggest that the observed
flattening of the PDMF towards lower masses is an intrinsic feature of the
LMC IMF. A more definitive statement regarding the LMC IMF, however, can
only be obtained once the LMC SFH has been derived in an unambiguous way.

As far as the younger stellar population, which is also located at this
part of the CMD, is concerned we compared the values of the initial
masses, corresponding to the same magnitudes in the old ($\tau$ \gsim\
10 Gyr), the intermediate (2 \lsim\ $\tau$/Gyr \lsim\ 10), and the
young (0.5 \lsim\ $\tau$/Gyr \lsim\ 2) isochrone models, and we
found that the larger differences in the initial mass of the faint
main-sequence stars appear toward the brighter limits.  These differences
do not account for more than $\sim$ 0.1 $M${\solar}, and they are
extremely small toward the fainter magnitudes. Consequently the most
age-sensitive mass bins of the constructed MF are the ones between 0.9 and
1 $M${\solar}.  Since the MF shown in Figure \ref{fig_wfmf} has a mean
mass bin width of 0.03 $M${\solar}, the 10\% differences in the mass
estimation of individual stars can account for up to three mass bins, and
consequently may change the MF slope as it is estimated, by redistributing
stars in neighboring bins.

In order to check if this is the case we constructed the MF of the area
using wider bins ($\sim$ 0.1 $M${\solar}), and we applied again a linear
regression to estimate the MF slopes in different mass intervals and to
check for any significant changes in the MF. Of course for this MF we were
able to verify these differences with coarser uncertainties, because of
the use of wider bins.  The $\chi^{2}$ test showed that indeed there is a
linear correlation between $\log{F(m)}$ and $\log{m}$, which appears for
the mass range from the higher mass bin ($\sim$ 2.4 $M${\solar})  down to
$\sim$ 1 $M${\solar}.  According to the same test there is no linear
correlation at all for wider mass ranges. The MF slopes were found to be
comparable to the ones estimated for the MF of narrower binning for almost
the same mass range.

In general, with the tests above and the construction of a MF with wider
bins we were able to verify that the use of a unique M/L relation does not
affect the estimated MF slopes significantly. Consequently, we can safely
conclude that {\em the coexistence of low-mass stars of different age at
the lower part of the main sequence does not affect the corresponding MF
slope up to $\sim$ 0.9 - 1 $M${\solar}}. Under these circumstances the MF
of the main-sequence stars, shown in Figure \ref{fig_wfmf}, represents
their IMF for masses up to this mass limit.

\begin{figure*}[t!]
\centerline{\hbox{
\psfig{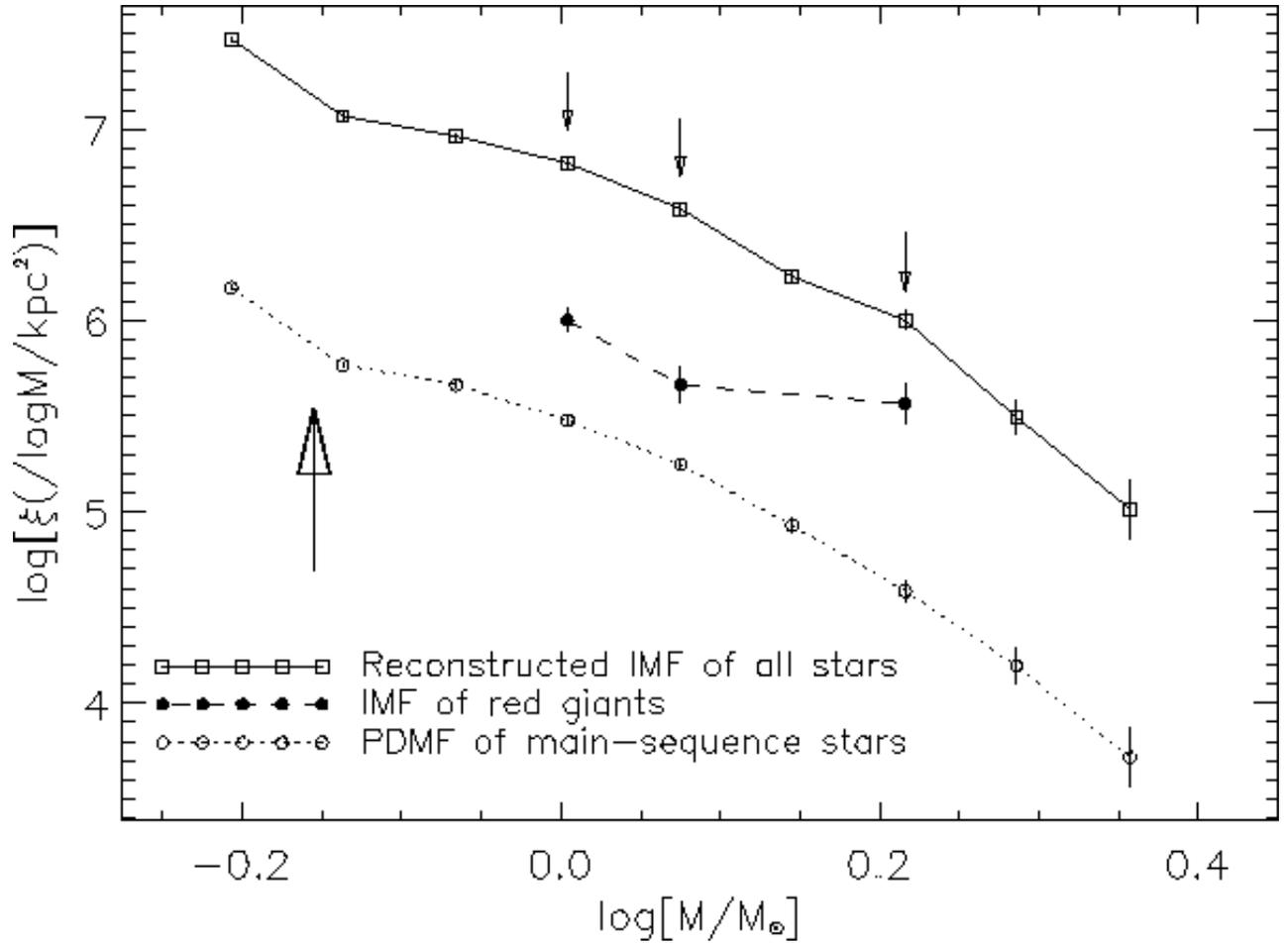}
}}
\caption{The Initial Mass Function of all stars in the observed area (red
giants and main-sequence), as it is reconstructed under the assumptions
discussed in \S 5, plotted with box symbols and the solid line. The small
number of RGB stars in the area and their narrow mass range introduce only
small changes in three specific mass intervals of the main-sequence MF
(small arrows), showing only a trend of the IMF to become slightly more
shallow toward the limit of $\sim$ 1.6 $M${\solar}. The red-giants IMF has
been constructed by counting stars in coarser mass intervals, because of
the small number statistics for the red giants, and it is plotted with
solid dots and dashed line. The main-sequence MF of Figure \ref{fig_wfmf}
constructed with the same binning is also plotted, with open dots and
dotted line. All mass functions shown are shifted to avoid overlapping.
The large arrow indicates the 95\% limit of completeness.}
\label{fig_wfimf}
\end{figure*}

\subsection{Assumptions concerning the larger masses}

As far as the upper main-sequence is concerned, i.e. main-sequence stars
with $M$ \gsim\ 0.9 $M{\solar}$, although their estimated masses are
almost equal to their initial masses according to the models, their
numbers per mass bin do not trace directly the IMF, because of the
existence of evolved stars, which should be taken into account. A smooth
and continuous star formation rate in the LMC (e.g. Smecker-Hane et al.\
2002) would naturally lead to a PDMF with a progressively steepening slope
towards higher masses. While all primordial stars with masses \lsim\ 0.9
$M${\solar} are still present on the main-sequence, the most massive stars
successively evolve off the main-sequence with increasing age of a stellar
population. As these stars ultimately end up either as black holes,
neutron stars, or white dwarfs, the optical data presented in this paper
could not account for them in any comprehensive way and in the following
discussion we consider only the red giants and white dwarfs (WDs).

Stars of mass comparable to that of the Sun evolve to form white dwarfs,
while above some critical mass, $M_{\rm c}$, they explode as Type II
supernovae instead. Predictions for $M_{\rm c}$ range from 6 to 10
$M${\solar}, depending on the models (Weidemann 1990;  Jeffries 1997;
Garc\'{i}a-Berro et al. 1997; Pols et al. 1998). Elson et al. (1998) have
identified a candidate luminous white dwarf with an age of 12 $\times$
$10^{5}$ yr in NGC 1818, a young star cluster in the LMC with HST/WFPC2
observations. This discovery constrains the boundary mass for WDs in the
LMC to $M_{\rm c}$ \gsim\ 7.6 $M${\solar}. Models also suggest that WDs
should have $(V-I)  \sim -0.4$ mag (e.g. Wood 1991; Cheselka et al. 1993),
a value which is roughly independent of age and metallicity. For the WFPC2
passbands $F555W$ and $F814W$ this color is equivalent to $(V-I)=-0.44$
mag (Holtzman et al. 1995), and thus the range of colors in which we might
expect to find WDs in the CMD of Figure \ref{fig_cmdiso} is $-0.64$ \lsim\
$(V-I)$/mag \lsim\ $-0.24$ (Elson et al. 1998). This gives roughly 25
candidate WDs (not corrected for completeness), almost half of which have
magnitudes $V$ \lsim\ 24.5 mag, while the rest are very close to the
detection limit and can be spurious detections as well. This very small
number leads to the conclusion that the contribution of WDs to the IMF of
our fields should be negligible.


Therefore, in order to correct the MF for evolutionary effects we consider
the existence of red giants only, and we make use of their initial mass
per magnitude provided by the Padova models, assuming that no nova or
super-nova explosion occurred in the observed field. A simple assumption
that can be made is that few old isochrones are sufficient for the
estimation of the masses of the observed RGB stars shown in the CMD of
Figure \ref{fig_cmdiso}. Although, the stellar masses according to the
isochrone models do not differ significantly from one model to the next on
the main-sequence, this is not the case for the red-giant branch.
Specifically, the mass provided by the 10 Gyr model for a typical LMC
metallicity (Z $=$ 0.008) for RGB stars with -1 \lsim\ $M_{\rm V}$
\lsim\ 4 is between 0.97 and 0.99 $M${\solar}, while the model for
$\sim$ 2 Gyr gives masses 1.59 \lsim\ $M/M${\solar} \lsim\ 1.66 for
the same magnitude range.  On the other hand metallicity does not seem to
affect significantly the resulting masses. For example, the corresponding
masses from models of the same age for lower metallicity (Z $=$ 0.0001)
are 0.86 - 0.87 for 10 Gyr and 1.42 - 1.44 for 2 Gyr. Consequently, in
order to establish a realistic representation of the IMF of our RGB stars,
a careful selection of the models for the estimation of their initial
masses should be made.

This selection is based on the results on the SFH of the LMC, discussed
earlier (\S 3.1). These results can be summarized as a major increase in
the SFR occurring about 2 - 4 Gyr ago (Elson et al. 1997; Castro et al.
2001) and according to other authors about 2 Gyr ago (Gallagher et al.
1996; Geha et al. 1998;  Smecker-Hane et al. 2002; Javiel et al. 2005).
Gallagher et al. (1996) suggest that this later enhancement in the SFR
resulted in 25\% of the field population, while Smecker-Hane et al. (2002)
found that the increase in the SFR 1 - 2 Gyr ago produced 15\% of the
stellar mass in the bar. Taking these numbers into account it would be
reasonable to assume that 20\% of the RGB population was formed around 2
Gyr ago. Furthermore, considering that 25\% of the stars in the bar were
probably formed about 4 - 6 Gyr ago (Smecker-Hane et al. 2002), we can
assume that 25\% of the RGB population in our CMD has an age of about 5
Gyr. Finally, since the LMC field population is dominated mostly by stars
of age $\sim$ 10 Gyr (Castro et al. 2001), the model of this age should be
used for the rest of our RGB stars. Ultimately, the Padova isochrone model
of $\sim$ 2 Gyr for 20\% of the RGB stellar population, the one of $\sim$
5 Gyr for 25\% of the population and the model for 10 Gyr for 55\% of the
population will be used for the estimation of their initial masses and the
construction of the red-giant IMF of our sample.

Taking all the above hypotheses into account for the reconstruction of the
IMF of the observed area, the resulting IMF may be considered as a rough
representation of the IMF of the LMC field at the edge of its bar, and as
the most realistic as possible, within the uncertainties caused by our
assumptions. According to the models used for the estimation of the masses
of the red giants in our sample, these stars fall in few very narrow mass
ranges, not allowing the construction of their IMF with the fine binning
of the main-sequence MF shown in Figure \ref{fig_wfmf}.  Thus, we counted
all red giants in wider logarithmic mass intervals, which resulted in only
three bins for their IMF, which are plotted in Figure \ref{fig_wfimf}
(thick dots - dashed line). We constructed the main-sequence MF with the
same grosser binning and we also show it in Figure \ref{fig_wfimf} for
comparison (open dots - dotted line). The small numbers of bins in the
red-giant IMF only provides a qualitative estimate of the IMF. 

From Figure \ref{fig_wfimf} it can be seen that the slope of the red-giant
IMF tends to be more shallow than the main-sequence MF for the same mass
range (1 \lsim\ $M/M${\solar} \lsim\ 1.6). This trend is also
present in the IMF of the whole population, which is shown as it was
constructed with the use of the same binning (Figure \ref{fig_wfimf}),
making it a bit more shallow for the same mass range (the affected bins
are indicated by the small arrows in the figure). Still, the small number
of RG in our sample compared to the MS population ($\sim$ 1900 red giants
to $\sim$ 75,000 stars selected as main-sequence stars), cannot change
significantly the slope of the MF already shown in Figure \ref{fig_wfmf}.
If this tentative result represents reality then {\em the IMF of the LMC
field in the observed area does not differ much from its present-day MF
for the mass range 0.7 \lsim\ $M$/$M${\solar} \lsim\ 2.4}.

\section{Conclusions and Discussion}

We made use of a large sample of almost 80,000 stars observed with
HST/WFPC2 in a region of about 32 arcmin$^{2}$ in the general field west
of the bar of the LMC to construct its main-sequence PDMF and, based on
several assumptions, to qualitatively reconstruct its IMF. The conclusions
of this study can be summarized as: (1) {\em The main-sequence LF of the
observed LMC field is in very good agreement with the Galactic LF} as it
was previously found (Kroupa et al. 1990) within the overlapping magnitude
range of 3 \lsim\ $V$/mag \lsim\ 6. (2) We verified statistically that in
the observed area {\em the main-sequence MF of the LMC field does not
follow a single power law}, but it changes at about 1.0 $M${\solar} to
being shallower for stars with smaller masses down to the lower observed
mass (within 95\% completeness) of about 0.7 $M${\solar}. (3) The
main-sequence PDMF of stars with masses between $\sim$ 0.9 $M${\solar} and
0.7 $M${\solar} is found well correlated with mass with a slope $\Gamma$
starting from $-$1.4, comparable to Salpeter's IMF, to become flatter for
shorter mass ranges (Table 3). This provides a clear hint of flattening of
the main-sequence MF of the LMC field below $\sim$ 0.7 $M${\solar}. This
MF accounts for the IMF of the LMC field in the low-mass regime. The IMF
becomes very flat with $\Gamma$ \simsim\ $-0.3$ for the lowest observed
masses (within the 95\% completeness), but this result is based only on
three mass bins, and thus suffers from poor statistics. Deeper
observations would certainly provide a larger number of stars in this mass
range and thus more information on the low-mass flattening of the LMC
field IMF. (4) The slope of the main-sequence MF becomes a bit steeper for
masses higher than $\sim$ 0.9 up to $\sim$ 1 $M${\solar} ($-$2.9 \lsim\
$\Gamma$ \lsim\ $-$2.7). For masses between $\sim$ 1 $M${\solar} and
the highest observed mass ($\sim$ 2.3 $M${\solar}) it is even steeper with
$-$5.2 \lsim\ $\Gamma$ \lsim\ $-$4.5, similar to the one previously
found in the same mass range (Gouliermis et al. 2005), and for massive
stars (Massey et al. 1995). (5) We attempted a qualitative reconstruction
of the IMF in the whole observed mass range, taking into account
evolutionary effects for the whole observed mass range 0.7 \lsim\
$M/M${\solar} \lsim\ 2.3.

A comparison between the IMF in clustered star-forming regions of low and
high density environments (e.g. Hill et al. 1994; Hill et al. 1995) shows
that there are systematic differences in the IMF from one stellar system
to the next in the LMC. However, Massey \& Hunter (1998) found that the
IMF slope in R136 in 30 Doradus is indistinguishable from those of
Galactic and Magellanic Cloud OB associations and they suggest that star
formation produces the same distribution of masses over a range of $\sim$
200 times in stellar density, from that of sparse OB associations to that
typical of globular clusters. Indeed the IMF slopes of LMC associations
are found by various authors (Massey et al. 1989a,b, 1995;  Parker et al.
1992; Garmany et al. 1994; Oey \& Massey 1995; Oey 1996;  Dolphin \&
Hunter 1998; Parker et al. 2001; Olsen et al. 2001; Gouliermis et al.
2002) are similar, considering the observational constraints,
and are clustered around $\Gamma$ $\simsim$ $-1.5 \pm 0.1$ for
intermediate-to-high-mass stars. This slope is not very different from
the mass function slopes of typical LMC clusters for the same mass range
(e.g. Hunter et al. 1997; Fischer et al. 1998; Grebel \& Chu 2000; de
Grijs et al. 2002; Gouliermis et al. 2004). In general, if the observed
differences from cluster to cluster are subject to systematic
uncertainties, or if the small fluctuations around a typical Salpeter
IMF are related to the density of the regions is still unknown.

On the other hand the observed difference between the LMC field high-mass
star IMF (e.g. Massey et al. 1995; Parker et al. 1998), which has a slope
around $\Gamma \simeq -4$, and the IMFs found for stellar associations in
this galaxy (see references above) is larger than the measured 
uncertainties, and thus it cannot be accounted entirely to observational 
constraints in the detection of low-mass stars. Furthermore, Gouliermis et
al. (2005) recently showed that the field of the LMC is characterized by  
the majority of the observed stars with $M < 2$ $M${\solar} with an IMF  
slope $\Gamma$ \simsim\ $-5$. Such a difference of the IMF slope has been
also observed in the solar neighborhood (Scalo 1986; Tsujimoto et al.    
1997), but in this case corrections for low and intermediate masses, such 
as the loss of evolved stars and possible variations in the past star 
formation rate should be taken into account (Scalo 1986). Consequently,
the solar neighborhood field may be considered as a composite of several
different cluster IMFs from aging dispersed loose stellar systems and thus
the corresponding IMF at intermediate and/or high mass could be steeper
than the IMF in a typical young cluster if the low mass stars
systematically drift further from their points of origin than higher mass
stars.

In the case of the LMC even though the field IMF is observed to be steep,
it could still originate from the same shallow IMF observed in stellar   
systems after they disperse. Differential evaporation of stars from the   
periphery of dispersed stellar systems or differential drift of long-lived
low-mass stars into the field are expected to steepen the field IMF (e.g.
Elmegreen 1997). An example of such a process has been presented by      
Gouliermis et al. (2002), who found a clear difference between the IMF   
slope of association LH 95 in the LMC, its surrounding field and the 
general field of the galaxy for the same mass range (3 \lsim\
$M$/$M${\solar} \lsim\ 10), with the IMF becoming gradually steeper  
outwards from the main body of the association. They interprete this
phenomenon as due to the evaporation of the dispersed association, which 
feeds the general LMC field with intermediate-mass stars through its     
surrounding field, while the system itself is characterized by a centrally
concentrated clump of massive stars, as if mass segregation takes place.
Intermediate-mass stars (\lsim\ 15 $M${\solar}) in the outer parts of 
mass segregated young LMC clusters have steep IMF with slopes $\Gamma$   
\lsim\ $-2$ (de Grijs et al. 2002a; Gouliermis et al. 2004). Whether    
this kind of stars in associations have the time to migrate and produce   
the observed steep field IMF is an open issue.

In any case Elmegreen (1999) notes that either the IMF is independent of  
star-forming density, and the general LMC field is somehow not a
representative sample, or there is a threshold low density where the IMF 
abruptly changes from Salpeter-like to something much steeper at lower   
density. He proposes a field IMF that is a superposition of IMFs from many
star-forming clouds, all with different masses, and he assumes that the   
largest stellar mass in each cloud increases with the cloud mass, perhaps
because it takes a more massive star to destroy a more massive cloud.  He
suggests that since many clouds, even small ones, can produce low mass   
stars, but only the larger clouds produce high mass stars, the summed IMF
will be steeper than the individual cloud IMF. However, his previously
developed model for the stellar IMF (Elmegreen 1997), which is based on
random selection of gas pieces in a hierarchical cloud, with a selection
probability proportional to the square root of density and a lower mass
cutoff from the lack of self-gravity, cannot explain the steep slope of
the extreme field IMF without considering {\em significantly different    
physical effects}.

More recently, Elmegreen (2004) introduced a multicomponent IMF model,
where he discusses three characteristic masses and their possible origins.
He presents two examples where different parts of the IMF are relatively  
independent to demonstrate that the observed power law distributions
ranging from solar-mass to high-mass stars {\em do not necessarily imply a
single scale-free star formation mechanism, but the IMF can be a composite
of IMFs from several different physical processes}.  Brown dwarfs with
masses of the order of $0.02$ $M${\solar} may be the result of dynamical
processes inside self-gravitating pre-stellar condensations or
gravitational collapse in the ultra high-pressure shocks between these 
condensations.  Solar-to-intermediate mass stars could be formed from the
pre-stellar condensations themselves, getting their characteristic mass
from the thermal Jeans mass in the cloud core.  High-mass stars could grow
from enhanced gas accretion and coalescence of pre-stellar condensations.
These processes, blended together or with poor sampling statistics, can
produce what appears to be a universal IMF. But when IMF is viewed with
good statistics (like in our case here) on small scales and for specific
mass ranges a diversity shows up, which can be interpreted as a variable  
IMF. According to Elmegreen (2004) these processes, which may as well
coexist for all three considered mass ranges, broaden each component in   
his IMF models into what was approximated as a log-normal.

New results from HST/WFPC2 observations on the low-mass MF in the LMC,
were recently provided by Gouliermis et al. (2005). Two areas in the LMC
were studied, one on the stellar association named LH 52 (Lucke \& Hodge
1970) and one on the background field of the close-by association LH 55.  
It was found that the low-mass MF slope of the field in the LMC, is
independent of the location (an empty general field or the area of a star
forming association) and it has a value between $-$4 \gsim\ $\Gamma$ 
\gsim\ $-$6 for stars with masses 1 \lsim\ $M/M${\solar} \lsim\ 
2. In Gouliermis et al. (2005) the data did not contain information on the 
MF slope for lower masses, while in the present study we are able to get 
this information and we find statistically significant evidence that the 
MF becomes shallower for $M$ \lsim\ 1 $M${\solar}.

The investigation on the general field of the LMC by Massey et al. (1995)
gave IMF slopes, for stars $M$ \gsim\ 2 $M${\solar}, which have more or
less the same slope as we found for 1 \lsim\ $M/M${\solar} \lsim\ 2. This
result clearly implies that the IMF slope of the LMC field is much steeper
than Salpeter's, for the whole so far observed mass range with M \gsim\
1 $M${\solar}.  Furthermore, Chabrier (2003) in an extensive review
derives MF slopes for the Galactic field. We find that the slopes of the
IMF of our observed LMC field, given in Table 2, are comparable to the
ones by Chabrier for stars with $M$ \gsim\ 1 $M${\solar}.

\acknowledgments

This paper is based on observations made with the NASA/ESA Hubble Space
Telescope, obtained from the data archive at the Space Telescope Science
Institute. STScI is operated by the Association of Universities for
Research in Astronomy, Inc. under NASA contract NAS 5-26555. This research
has made use of NASA's Astrophysics Data System Bibliographic Services
(ADS), of the SIMBAD database, operated at CDS, Strasbourg, France, and of
{\em Aladin} (Bonnarel et al. 2000).





\end{document}